\documentclass[%
 reprint,
superscriptaddress,
nofootinbib,
 amsmath,amssymb,
 aps,
]{revtex4-2}
\usepackage{graphicx}
\usepackage{dcolumn}
\usepackage{bm}
\usepackage[version=4]{mhchem} 
\usepackage{subfig}
\usepackage{float}
\usepackage{caption}
\usepackage{multirow}
\usepackage[colorlinks,
            linkcolor=red,
            anchorcolor=blue,
            citecolor=blue
           ]{hyperref}

\graphicspath{ {figure/} }
\begin{document}

\title{Possible spectral irregularities in the AMS-02 positron spectrum}

\author{Xing-Jian Lv}
\email{lvxj@ihep.ac.cn}
 \affiliation{%
 Key Laboratory of Particle Astrophysics, Institute of High Energy Physics, Chinese Academy of Sciences, Beijing 100049, China}
\affiliation{
 University of Chinese Academy of Sciences, Beijing 100049, China 
}%
 \author{Xiao-Jun Bi}
 \email{bixj@ihep.ac.cn}
\affiliation{%
 Key Laboratory of Particle Astrophysics, Institute of High Energy Physics, Chinese Academy of Sciences, Beijing 100049, China}
\affiliation{
 University of Chinese Academy of Sciences, Beijing 100049, China 
}%
\author{Kun Fang}
\email{fangkun@ihep.ac.cn}
\affiliation{%
 Key Laboratory of Particle Astrophysics, Institute of High Energy Physics, Chinese Academy of Sciences, Beijing 100049, China}
 \author{Peng-Fei Yin}
\email{yinpf@ihep.ac.cn}
\affiliation{%
 Key Laboratory of Particle Astrophysics, Institute of High Energy Physics, Chinese Academy of Sciences, Beijing 100049, China}
\author{Meng-Jie Zhao}
\email{zhaomj@ihep.ac.cn}
 \affiliation{%
 Key Laboratory of Particle Astrophysics, Institute of High Energy Physics, Chinese Academy of Sciences, Beijing 100049, China}
\affiliation{
China Center of Advanced Science and Technology, Beijing 100190, China 
}%



\date{\today}

\begin{abstract}
The excesses in the electron and positron spectra observed by many experiments, such as PAMELA and AMS-02, have sparked significant theoretical investigation. 
It is not easy to distinguish the two primary hypotheses dark matter annihilation/decay and pulsars from the spectral features.
Should pulsars be the source of this excess, the expected variability in their distribution may introduce distinct irregularities in the positron energy spectrum. In this study, we use an irregularity estimator to detect these potential features in the positron energy spectrum of AMS-02. Our analysis of the current AMS-02 data reveals these spectral irregularities with a statistical significance of $1.75\sigma$. However, our projection indicates that, with AMS-02 data collected over a period of 20 years, such irregularities could be identified with a confidence level of $3\sigma$ level in 71\% of our simulations.
\end{abstract}
\maketitle


\section{\label{sec:level1}INTRODUCTION}
Astrophysical evidence overwhelmingly supports the existence of dark matter (DM), yet the non-gravitational detection of DM particles remains elusive. The measurements of PAMELA~\cite{PAMELA:2008gwm} and AMS-02~\cite{AMS:2019rhg} have reported unexpected excesses in the cosmic ray (CR) positron flux above 10 GeV. These anomalous positron fluxes may originate from DM annihilation or decay within the Galactic halo, e.g.~\cite{Bergstrom:2008gr, Barger:2008su, Cirelli:2008pk, Yin:2008bs, Arkani-Hamed:2008hhe, Pospelov:2008jd, Cholis_2009,Hamaguchi_2009, Lin:2014vja, Wang:2018pcc, Cheng:2016slx}, or be attributable to nearby astrophysical sources such as pulsars, e.g.~\cite{aharonianHighEnergyElectrons1995, Atoian:1995ux, Kobayashi:2003kp, Profumo:2008ms, Hooper:2008kg, Yuksel:2008rf, Malyshev:2009tw, Blasi:2009hv, Ioka_2010, Yin:2013vaa, DiMauro:2015jxa, Cholis:2022kio}. Notably, high-energy electrons and positrons undergo substantial energy losses through synchrotron radiation and inverse Compton scattering.
The relationship between the maximum energy that electrons and positrons can possess and the source distance $R$ could be estimated by \( E_{\text{max}} \approx 100 \, \text{GeV} \times (R/2 \, \text{kpc})^{-2} \) \cite{Cholis:2017ccs}. Therefore, any potential sources of the positron excess should be nearby, confined to a local volume of approximately $\sim \text{kpc}^3$. The primary objective of this study is to identify an observational signature that could decisively differentiate between these two plausible theoretical propositions.

Pulsars (or pulsar wind nebulae) are established astrophysical accelerators of high-energy electrons and positrons~\cite{1974MNRAS1671R}, as evidenced by multi-wavelength observations~\cite{Slane_2017}.
The discovery of high energy pulsar gamma-ray halos further indicates that the accelerated electrons and positrons can escape from the pulsar wind nebulae and diffuse into the interstellar medium~\cite{HAWC:2017kbo,LHAASO:2021crt,Fang:2022qaf,HAWC:2023jsq}. The Milky Way gives rise to pulsars at an approximate rate of one per century~\cite{1999MNRAS302693D,Vranesevic:2003tp,Faucher-Giguere:2005dxp,Keane:2008jj,Lorimer:2006qs}, and these newly formed pulsars are theorized to impart a significant portion of their rotational energy into electrons and positrons in the initial phase of their life cycle~\cite{Malyshev:2009tw, Profumo:2008ms, Cholis:2021kqk}.
\begin{figure}[htbp]
\includegraphics[width=0.48\textwidth]{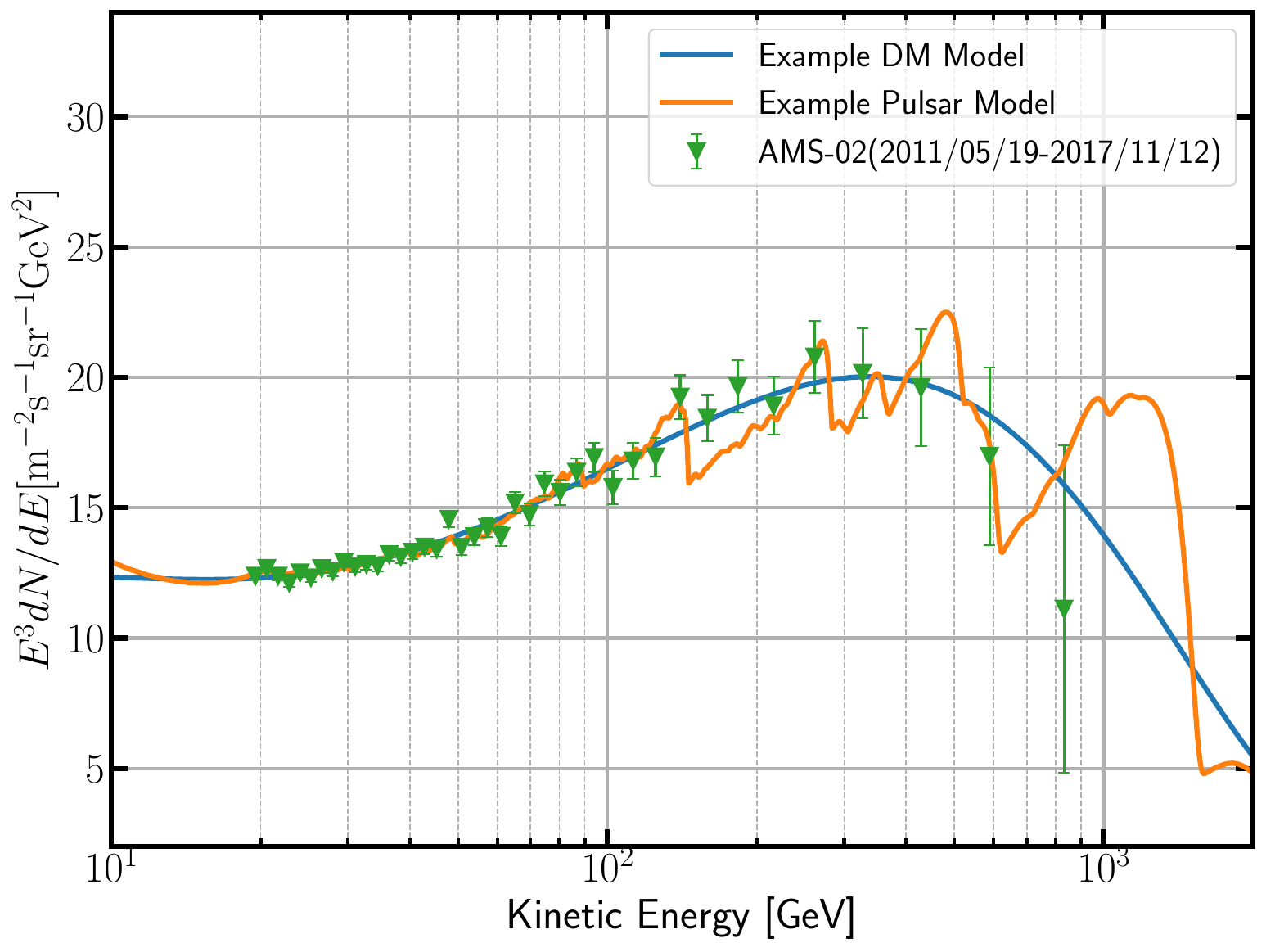}\\
\captionsetup{justification=raggedright}
\caption{The AMS-02 positron measurement~\cite{AMS:2019rhg} and two examples of models that fit it well. The orange curve illustrates contributions from a collection of Milky Way pulsars taken from Ref.~\cite{Cholis:2021kqk}. The blue curve represents a smooth fit to the AMS-02 data~\cite{AMS:2019rhg}, which could be interpreted as the contribution from DM annihilation or decay. Notably, the DM spectrum manifests as inherently smooth, in contrast to the pulsar spectrum, which reveals discernible contributions from discrete sources. 
\label{fig:schematic}}
\end{figure}
Due to the necessity for any additional sources explaining the positron excess to be nearby, only a limited number of pulsars are in a position to contribute to the positron flux above 10 GeV. This is particularly true when accounting for the slow-diffusion zone surrounding pulsars \cite{Fang:2019ayz}, the scale of which could span tens of parsecs \cite{Fang:2021qon,2023arXiv231016594F}. This scarcity of contributing pulsars may impart discrete features within the CR positron spectra~\cite{Profumo:2008ms, Fermi-LAT:2009ppq, Malyshev:2009tw, Yin:2013vaa, Cholis:2017ccs,Fornieri:2019ddi}, as schematically depicted in Fig.~\ref{fig:schematic}. Notably, these characteristics are unique and not reproducible by DM scenarios, even if multiple DM particles are considered~\cite{Cholis:2009va,Dienes:2013lxa}.

In this work, we use an irregularity estimator to search for the spectral features in the positron spectrum measured by the AMS-02 experiment. This method has been used for investigating the irregularities in gamma-ray spectra induced by the axion-photon oscillation effect in Ref.~\cite{HESS:2013udx}. We find that although the current AMS-02 data only shows less than 2$\sigma$ deviation from a smooth spectrum, by 2030, most of our data-driven simulations show a 3$\sigma$ deviation from a smooth spectrum. We substantiate this prediction using a mock spectrum derived from a hypothetical population of nearby pulsars in Ref.~\cite{Cholis:2021kqk}. 

This work is organized as follows. In Section~\ref{sec methology}, we describe 
the specific methodologies and datasets used in this analysis. Our results are presented in section~\ref{sec results}. In section~\ref{sec:conclusion}, we give a summary of our findings.

\section{Method}\label{sec methology}
\subsection{Data}
In this study, we use the published AMS-02 positron data~\cite{AMS:2019rhg} collected over 6.5 years, focus on the energy range above $20$ GeV, where the exotic component becomes dominate~\cite{AMS:2019rhg}. Our analysis includes only the statistical and unfolding errors, as other forms of systematic errors are deemed unlikely to mimic the spectral irregularities under examination~\cite{Cholis:2022kio}.
Furthermore, we do not incorporate the correlation between unfolding errors into our analysis, as they are subdominate at all energies.

For predictive purposes, we generate $10^3$ mock data sets after 20 years of data taking by AMS-02. The central values for these simulated datasets are extracted from a Gaussian distribution, centered on the current AMS-02 measurements and the contemporary statistical uncertainties serving as the standard deviation. While maintaining the initial unfolding error constant, we scale down the statistical error by a factor of $\sqrt{3}$.
This approach assumes that the present-day AMS-02 data are an accurate reflection of the true CR flux and that any deviations are primarily due to statistical fluctuations. We intentionally omit the data point above 1 TeV which is expected to be within the reach of AMS-02 over 20 years, owing to its indeterminate central value and the high probability of it being influenced by secondary processes\cite{AMS-02-ICRC2023}.
\subsection{Technique}
\begin{figure}[htbp]
\includegraphics[width=0.48\textwidth]{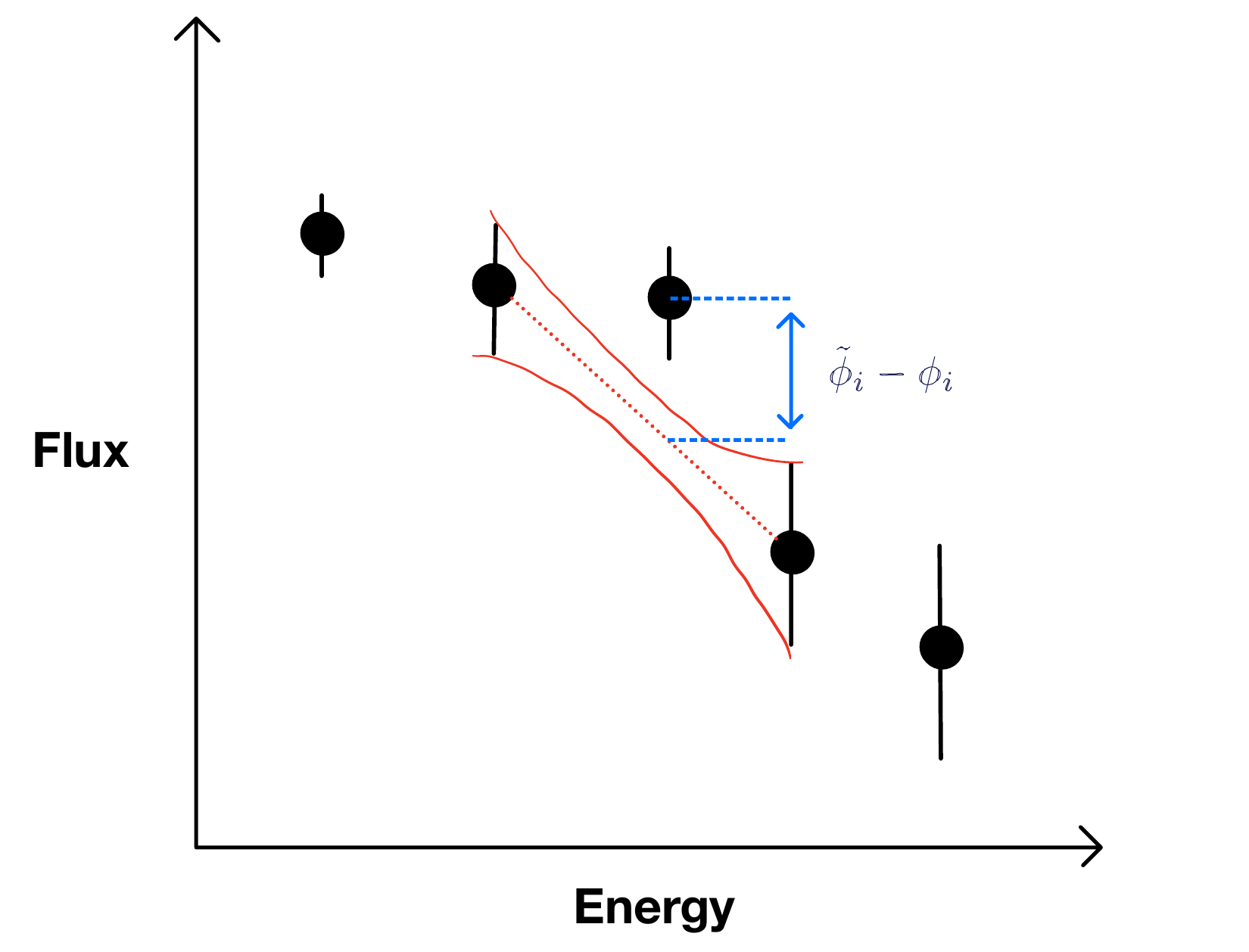}\\
\captionsetup{justification=raggedright}
\caption{Schematic picture for the residual $\tilde{\phi}_i - \phi_i $ accouting for irregularity.\label{fig:hess}}
\end{figure}

In this study, we implement the technique used by the H.E.S.S. collaboration in their investigation of potential spectral irregularities due to axions in the gamma-ray spectrum of PKS 2155-304~\cite{HESS:2013udx}.
The underlying hypothesis is that in the absence of contributions from nearby pulsars, the positron spectrum can be locally approximated by a power-law function. To test the local power-law behavior, we examine the energy ranges of three consecutive bins in the spectrum, as illustrated in Fig.~\ref{fig:hess}. Within such narrow energy intervals, deviations from a power-law behavior are not anticipated under the frameworks of secondary positron production or DM annihilation/decay. 
 
We segment the spectrum into independent sets of three consecutive bins, and generate 38 triplets for 40 energy bins. For the $i$th bin, $\tilde{\phi}_i$ is derived from the observed fluxes in the two surrounding bins $\phi_{i-1}$ and $\phi_{i+1}$, based on the power-law interpolation
\begin{equation}
\tilde{\phi}_i=\frac{\phi_{i+1}^{\beta_i}}{\phi_{i-1}^{\beta_i-1}} \text { with } \beta_i=\frac{\log \frac{E_{i-1}}{E_i}}{\log \frac{E_{i-1}}{E_{i+1}}} \;.
\end{equation}
For each triplet,  the residual $\tilde{\phi}_i - \phi_i $ is calculated as depicted in Fig.~\ref{fig:hess}. These residuals are then standardized by the associated uncertainties and summed quadratically to construct the irregularity estimator $ \mathcal{I}^2 $ ~\cite{HESS:2013udx}
\begin{equation}
\mathcal{I}^2=
\sum_i \frac{\left(\tilde{\phi}_i-\phi_i\right)^2}{
\left(\frac{\partial \tilde{\phi}_i}{\partial \phi_{i-1}}  \right)^2 \sigma_{i-1}^2
+\sigma_i^2+
\left( \frac{\partial \tilde{\phi}_i}{\partial \phi_{i+1}} \right)^2 \sigma_{i+1}^2} \;,
\label{eq:estimator}
\end{equation}
where $\sigma_i^2$ is the quadratic sum of the statistical error and unfolding error in the $i$th bin. As the correlated errors of the AMS-02 positron results are not published, they are not considered in the estimator. We also investigate irregularities spanning multiple bins, as discussed in Appendix~\ref{app: multiple}.

In the absence of spectral anomalies, the expected mean of \( \mathcal{I}^2 \) should equate to the number of triplets that can be formed. To establish the distribution of the irregularity estimator expected from multiple observational realizations devoid of nearby pulsar effects, we generate $ 10^4 $ random simulations based on the smooth spectral form initially proposed by Ref.~\cite{Cavasonza:2016qem} and later employed by the AMS-02 collaboration~\cite{AMS:2019rhg}
\begin{equation}
\begin{aligned}
\Phi_{e^{+}}(E)= & \frac{E^2}{\hat{E}^2}\left[C_d\left(\hat{E} / E_1\right)^{\gamma_d}\right. \\
& \left.+C_s\left(\hat{E} / E_2\right)^{\gamma_s} \exp \left(-\hat{E} / E_s\right)\right]
\end{aligned}
\label{eq:smooth}
\end{equation}
where $\hat{E} = E + \phi_{e^+}$. The free parameters $\{C_d, C_s, \gamma_d, \gamma_s, E_1, E_2, E_s, \phi_{e^+}\}$ are fitted by AMS-02 to the full positron spectrum. Our study, however, focuses on energies above 20 GeV, necessitating a refit of these parameters to improve the fit to the relevant part of the spectrum. In Table~\ref{table:refit}, we give the refitted values of the parameters in Eq.~(\ref{eq:smooth}) of  AMS-02 and our alternative parameterizations. As can be seen, the refitting process results in minor adjustments to certain parameters. 
We confirm that these adjustments do not materially affect our conclusions, as both the original and adjusted parameter sets yield similar $\mathcal{I}^2$ distributions. When generating the $\mathcal{I}^2$ distribution under the smooth spectrum assumption using the 20-year sample size of AMS-02, we also take the spectral form of Eq.~(\ref{eq:smooth}) and the parameters listed in Table~\ref{table:refit}.

\begin{table}[h!]
\centering
\begin{tabular}{lcc}
\hline
Parameter & AMS-02 value & Alternative value \\
\hline
$\phi_{e^+}$ (GeV) & 1.10 & 0.87 \\
$C_d$ (\( \text{m}^2 \text{sr GeV}^{-1} \)) & \( 6.51 \times 10^{-2} \) & \( 7.42 \times 10^{-2} \) \\
$E_1$ (GeV) & 7.0 & 70 \\
$\gamma_d$ & -4.07 & -4.05 \\
$C_s$ (\( \text{m}^2 \text{sr GeV}^{-1} \)) & \( 6.80 \times 10^{-5} \) & \( 6.74 \times 10^{-5} \) \\
$E_2$ (GeV) & 60.0 & 60.0 \\
$\gamma_2$ & -2.58 & -2.49 \\
$E_s$ (GeV) & 810 & 591 \\
\hline
\end{tabular}
\captionsetup{justification=raggedright}
\caption{The parameters describing the smooth function for the CR positron flux in Eq.~(\ref{eq:smooth}).}
\label{table:refit}
\end{table}

The simulated irregularity estimator distribution forms a probability density function (PDF) for the dark matter scenario. Using this PDF, the irregularity estimator calculated from actual AMS-02 data can be used to compute a p-value against the null hypothesis, which asserts the absence of spectral anomalies in the positron spectrum. 

For validation, we apply our methodology to the electron and antiproton spectra above 20 GeV as reported by AMS-02~\cite{AMS:2019iwo, AMS:2021nhj}. Given that these spectra\footnote{
High energy electrons also receive contributions from nearby sources. However, the primary sources of electrons are supernova remnants. These remnants, in contrast to pulsars, do not manifest as distinct, sharp features in the energy spectrum.} are not expected to manifest significant spectral anomalies, they act as control groups, corroborating the validity of our analytical technique. 

\section{RESULTS \& DISCUSSION}\label{sec results}

\begin{figure}[H]
\includegraphics[width=0.48\textwidth]{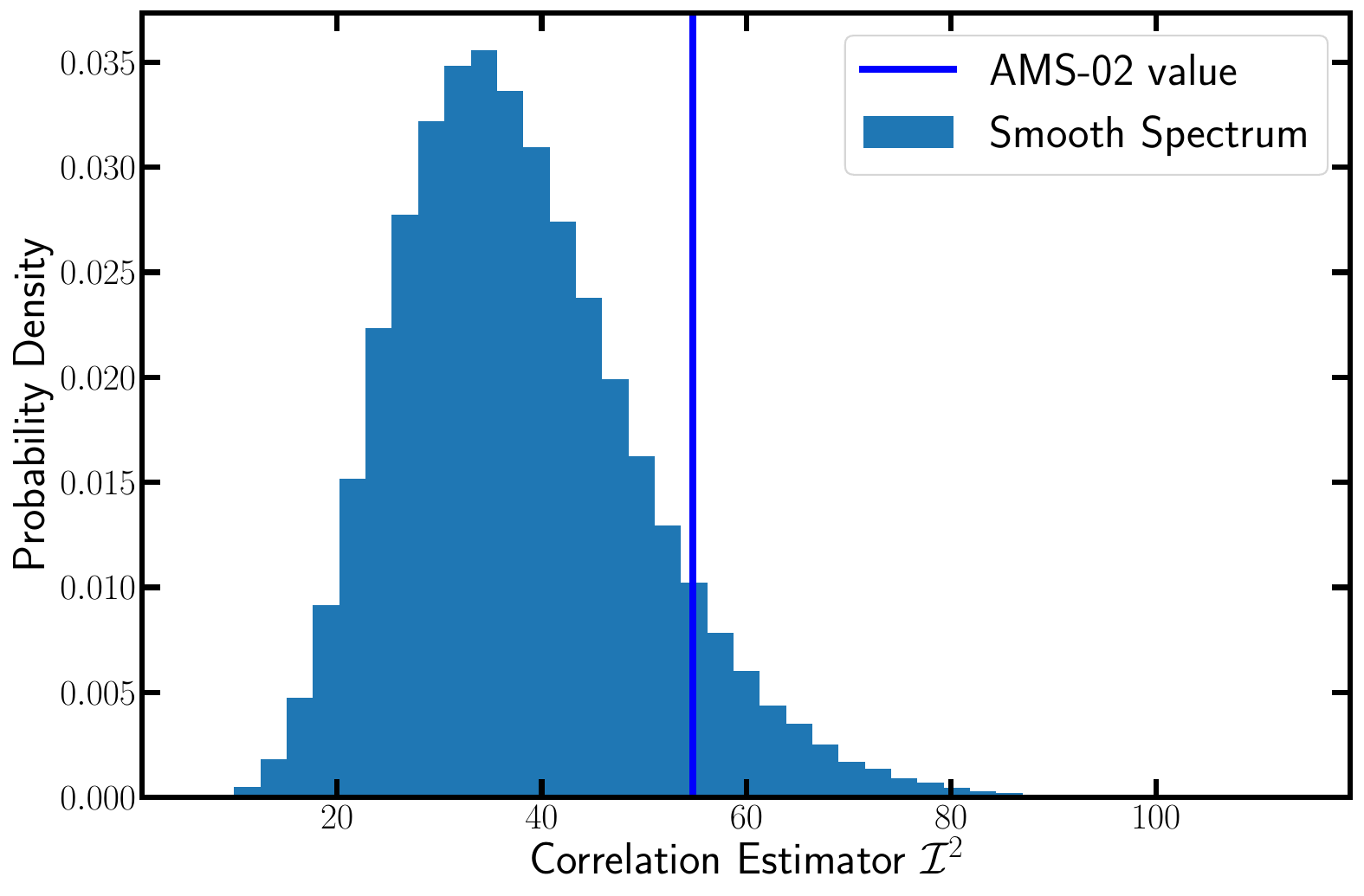}\\
\captionsetup{justification=raggedright}
\caption{Predicted PDFs for the irregularity estimator for positron spectrum under the assumption of a smooth spectrum. The vertical line represents the value of the irregularity estimator calculated from the AMS-02 measurement. \label{fig:irreg}}
\end{figure}

Our results are shown in Fig.~\ref{fig:irreg}, which presents the normalized distribution of the irregularity estimator for a theoretical smooth spectrum alongside the actual value derived from the AMS-02 data. As anticipated, the mean value of $\mathcal{I}^2$ aligns closely with $38$, matching the total number of triplets formulable from $40$ energy bins. A vertical line is drawn at $\mathcal{I}^2_{\text{AMS-02}} = 55$.  The calculated p-value for the null hypothesis,  which advocates for a smooth spectrum, stands at $0.08$. This result challenges the hypothesis of a smooth extra component in the spectrum with a statistical significance of $1.75\sigma$. This finding, while not disproving the smooth nature of the extra positron component, nevertheless casts doubt on it, and suggests that further scrutiny into the origin of the positron excess is warranted.

\begin{figure}[htbp]
\includegraphics[width=0.48\textwidth]{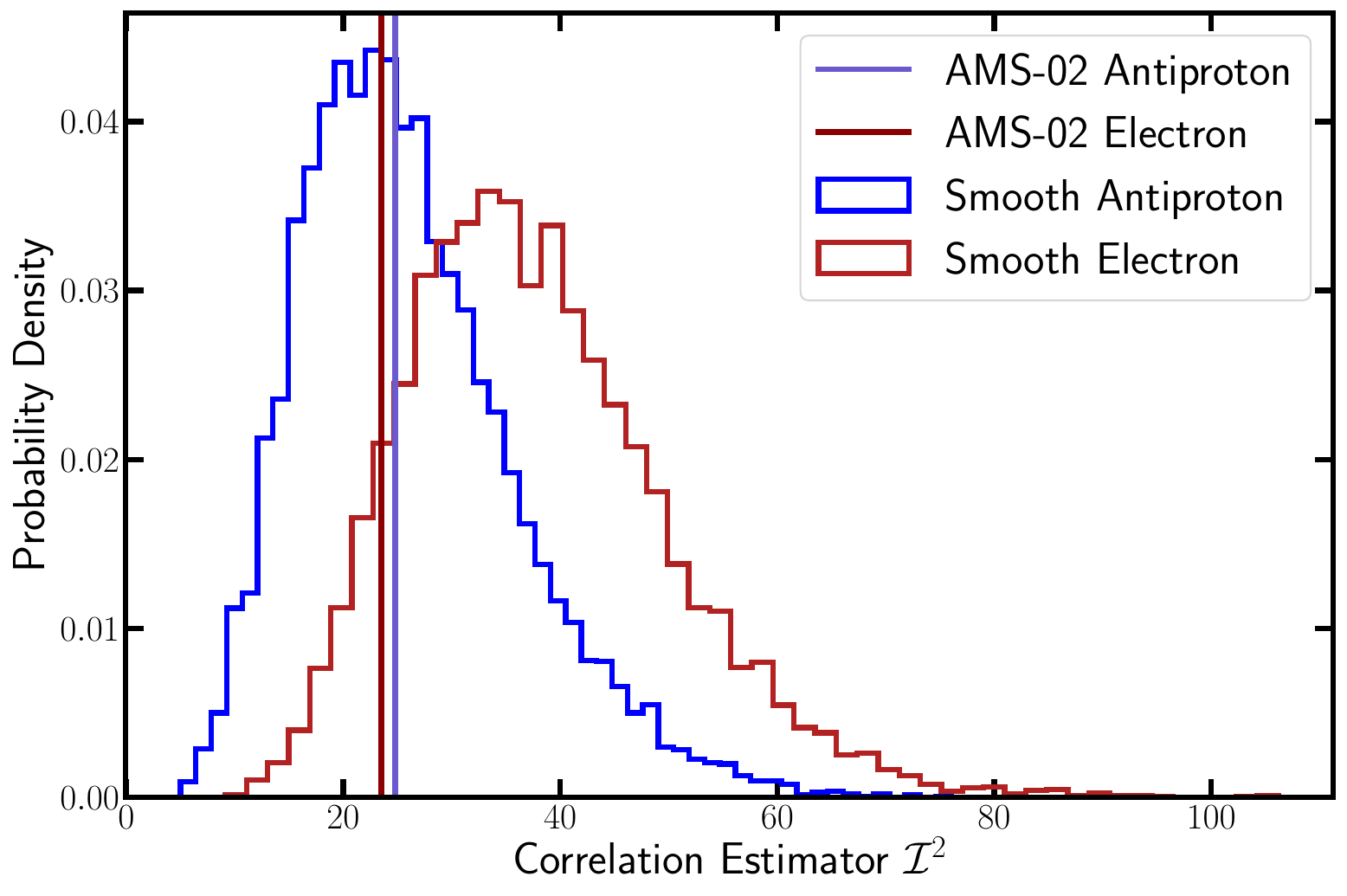}\\
\captionsetup{justification=raggedright}
\caption{The same as Fig.~\ref{fig:irreg}, but for the AMS-02 electron and antiproton data. \label{fig:sanity}}
\end{figure}

In Fig.~\ref{fig:sanity}, we apply the same analytical approach to the electron and antiproton spectra data from AMS-02. Consistent with expectations, the analysis yields no indication of spectral anomalies in either the electron or antiproton measurements. Specifically, the calculated irregularity estimators for the electron and antiproton spectra are 23.5 and 24.8, against 38 and 26 degrees of freedom, respectively. These estimators correspond to p-values of 0.9 for the electron spectrum and 0.5 for the antiproton spectrum, indicating a high probability that any observed fluctuations are consistent with statistical variation rather than indicative of true spectral features.
\begin{figure}[htbp]
\includegraphics[width=0.48\textwidth]{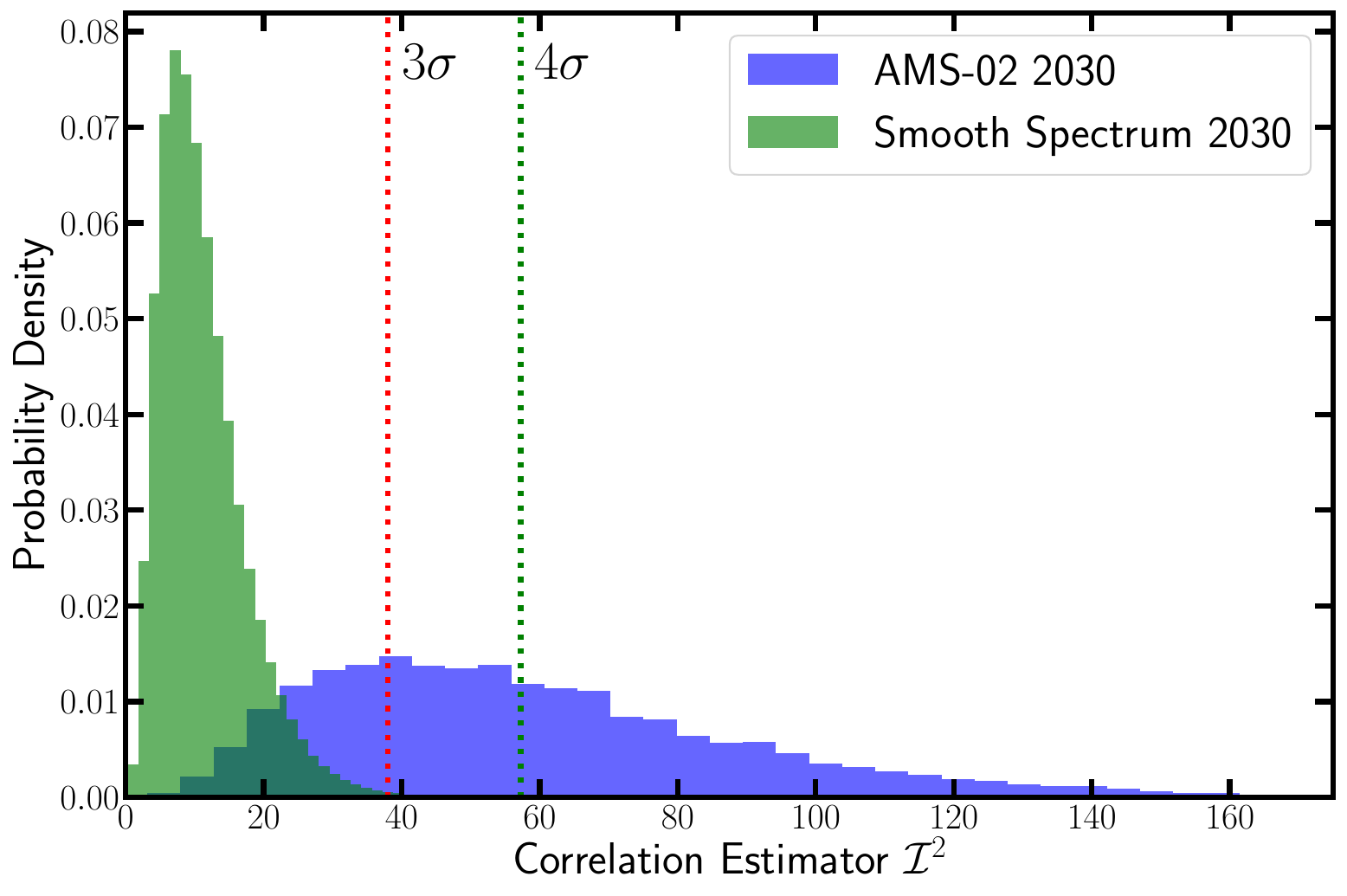}\\
\captionsetup{justification=raggedright}
\caption{Projected PDFs for the irregularity estimator of the positron spectrum above 100 GeV, derived from the anticipated AMS-02 sample size with 20 years of observation. The green histogram illustrates the smooth spectrum, while the blue histogram represents the anticipated AMS-02 measured value. The red and blue dotted lines indicate deviations at $3\sigma$ and $4\sigma$ levels, respectively, from the smooth spectrum.\label{fig:above100}}
\end{figure}

Given the current precision of the current AMS-02 data, it is insufficient to conclusively dismiss the hypothesis that dark matter contributes to the observed positron excess. Anticipating further data collection, we expand the projection to encompass a 20-year sample size for AMS-02, which is roughly three times of the size of the current dataset. As a result, we reduce statistical errors by a factor of $\sqrt{3}$, and assume consistent unfolding errors. Furthermore, it is viable to improve statistics through modified cut conditions~\cite{ams_cut}. Although this method may introduce an overall normalization error, it would not impact the detection of irregularities.  Therefore, reducing statistical error by $\sqrt{3}$ is a conservative choice. For comparison, a set of simulations based on the smooth spectrum is also performed. Our analysis is confined to energy points exceeding 100 GeV, a region where numerous pulsar models predict pronounced irregularities. As depicted in Figure \ref{fig:above100}, the projections are quite encouraging: a substantial 71\% of the simulated projections yield a $3\sigma$ deviation from those based on 
the smooth spectral distribution, while 44\% indicate an even more significant $4\sigma$ deviation. These findings underscore the potential of extended AMS-02 observations to elucidate the origins of the positron excess.

\begin{figure}[H]
\includegraphics[width=0.48\textwidth]{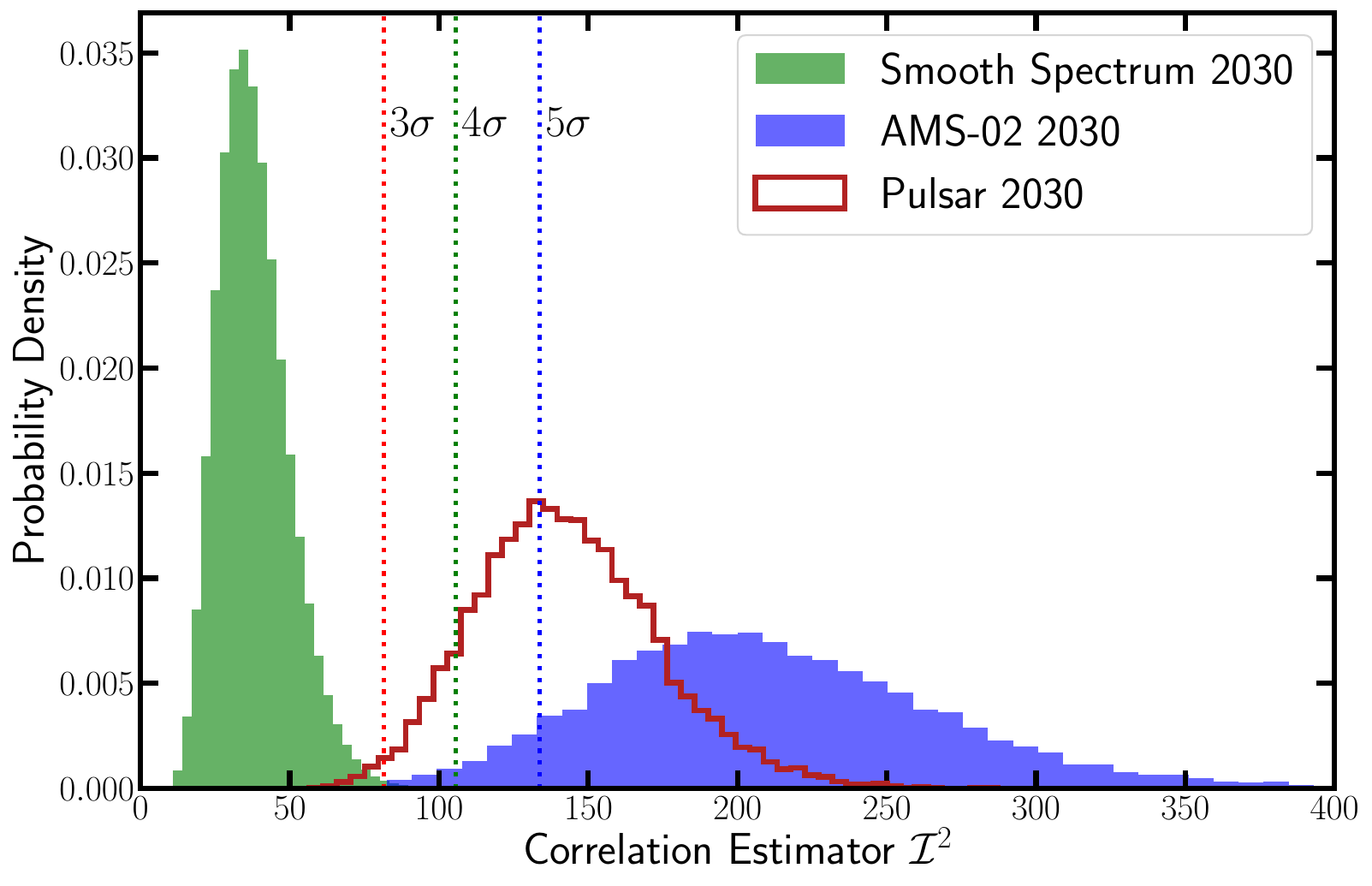}\\
\captionsetup{justification=raggedright}
\caption{The same as Fig.~\ref{fig:above100}, restricted to energies above 20 GeV. The vertical line denotes the contemporary value of $\mathcal{I}^2_{\text{AMS-02}}$ and the unfilled histogram is obtained from the mock-data generated by a representative pulsar model from Ref.~\cite{Cholis:2021kqk}.\label{fig:porjData}}
\end{figure}

Expanding the scope, we incorporate all data points above 20 GeV, with the results presented in Fig.~\ref{fig:porjData}.  A considerable portion of the simulated data sets manifest marked discrepancies from a smooth spectrum. More precisely, 99.8\% of the mock data deviates from a smooth spectrum with a significance level of $3\sigma$, 98.6\% at $4\sigma$, and 92.8\% at $5\sigma$. It is noteworthy that the inclusion of low-energy data increases the level of irregularity, and we further elaborate on this observation in Appendix~\ref{app: above100Gev}.

In the above analysis, the observed spectrum of AMS-02 with a 20-year sample size is assumed not to significantly deviate from the current result. This may not be the case if the current data accidentally contains some fluctuations. In Fig.~\ref{fig:porjData}, we also present the distribution of a spectral irregularity estimator based on a hypothetical group of nearby pulsars combined with secondary positrons, following a random model in Ref.~\cite{Cholis:2021kqk}\footnote{The calculations in Ref.~\cite{Cholis:2021kqk} treat the energy losses of the positrons as a continuous process. Contrastingly, Ref.~\cite{John:2022asa} highlights the importance of accounting for their stochastic nature, which can blur pulsar characteristics. However, the present work focuses on the possible spectral irregularities present in the AMS-02 data, thus exploring this aspect of variability is beyond our current research scope.}. This model also shows considerable deviation from a smooth spectrum, suggesting that the superposition of nearby pulsars could plausibly account for the mock spectral irregularities, projected up to the year 2030. 

Interestingly, the irregularities of this pulsar population appear to be less pronounced than those inferred from the AMS-02 data. This outcome underscores the incomplete nature of our understanding of positron injection and propagation from pulsars. Moreover, the inherent stochastic nature of the pulsar population compounds our lack of knowledge. For instance, the potential for multiple pulsars to contribute to the same energy range or for a nearby pulsar to significantly influence irregularities further complicates the analysis.
Importantly, the intent behind our analysis is not to favor any particular model of pulsar characteristics. Owing to the lack of precise information regarding the distribution of properties such as age and injection spectrum among local pulsars, our analysis is fundamentally aimed at illustrating the capacity of pulsars to produce significant spectral irregularities, rather than pinpointing the attributes of specific pulsar models. The complexity and uncertainty of pulsar models necessitate further research, with observed irregularities serving as valuable indicators for exploring pulsar population properties in greater depth.

\section{SUMMARY\label{sec:conclusion}}
In this study, we have outlined a statistical method utilizing an irregularity estimator to detect possible non-uniformities in the CR positron energy spectrum that could result from the discrete nature of pulsar sources, should they be the origin of the observed positron excess. Our findings indicate that the current dataset does not reveal irregularities of a magnitude that would allow for definitive conclusions. Nonetheless, projecting from the current central values obtained from AMS-02 data, our analysis suggests that such spectral irregularities would become significantly more apparent after two decades of data accumulation by AMS-02. This would potentially bring a resolution to the ongoing debate regarding the source of the positron excess.

In this analysis, we have processed binned data, but we acknowledge that leveraging raw, unbinned data might provide a more refined analysis, particularly in instances where the selected binning exceeds the instrumental resolution. Furthermore, incorporating correlations between systematic errors would enhance the robustness of the analysis. Therefore, we hope that the current work will act as a catalyst, encouraging the collaborations to undertake more detailed investigations into spectral irregularities using their original data sets.

\acknowledgments
This work is supported by 
the National Key R\&D Program Grants of China under Grant No. 2022YFA1604802, the National Natural Science Foundation of China under
the Grants No. 12175248 and No. 12105292. 

\bibliography{apssamp}
\appendix
\section{Test irregularities for multiple bins.}\label{app: multiple}

The individual contributions of single pulsars to the positron flux can extend across multiple energy bins, and the overlapping contributions from multiple pulsars could generate broader features. Therefore, we extend our method to detect irregularities spanning multiple bins. The generalization of our approach is straightforward: for $n$ consecutive bins, we employ the two endpoints to fit a straight line in log-log space and then calculate the sum of deviations from this line for the middle $n-2$ points. Notably, the analysis described in the main text for triplets corresponds to $n=3$.

\begin{figure}[htbp]
\includegraphics[width=0.48\textwidth]{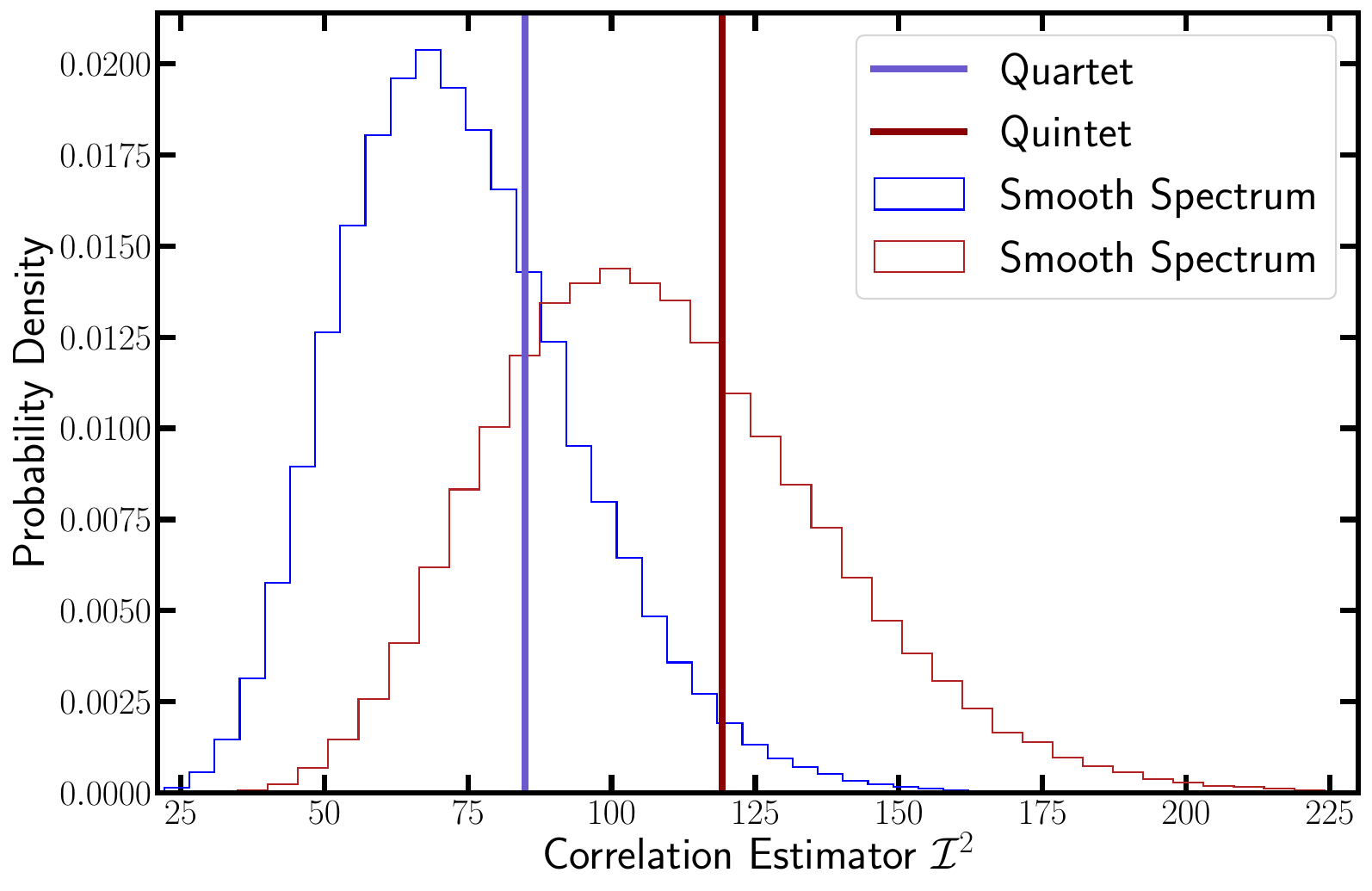}\\
\captionsetup{justification=raggedright}
\caption{The same as Fig.~\ref{fig:irreg}, but for quartet and quintet bins.}.\label{fig: multiBins}
\end{figure}

The results for $n=4$ and $n=5$ are shown in Fig.~\ref{fig: multiBins} for energies above 20 GeV. It is evident that the deviation from a smooth spectrum is less pronounced when examining sets of four and five bins in comparison to sets of three bins. The corresponding $p$-values for sets of four and five bins are 0.27 and 0.32, respectively. We do not use more bins, since contributions from dark matter annihilation/decay can also yield a broad spectral structure, as depicted in Fig.\ref{fig:schematic}.

Given that Fig.~\ref{fig:porjData} shows the AMS-02 data exhibits more irregularities than the pulsar model predicts and that the level of irregularities reduces when using more bins, it is likely that the AMS-02 data aligns more closely with the pulsar model when considering irregularities spanning multiple bins. However, the primary aim of this work is not to directly fit the irregularities estimator of the pulsar model with that of the AMS-02 data, and the uncertainties in pulsar modeling are substantial. A thorough consideration of these uncertainties is deferred to future work.

\section{Contributions to irregularities at different energies.}\label{app: above100Gev}

\begin{figure}[htbp]
\includegraphics[width=0.48\textwidth]{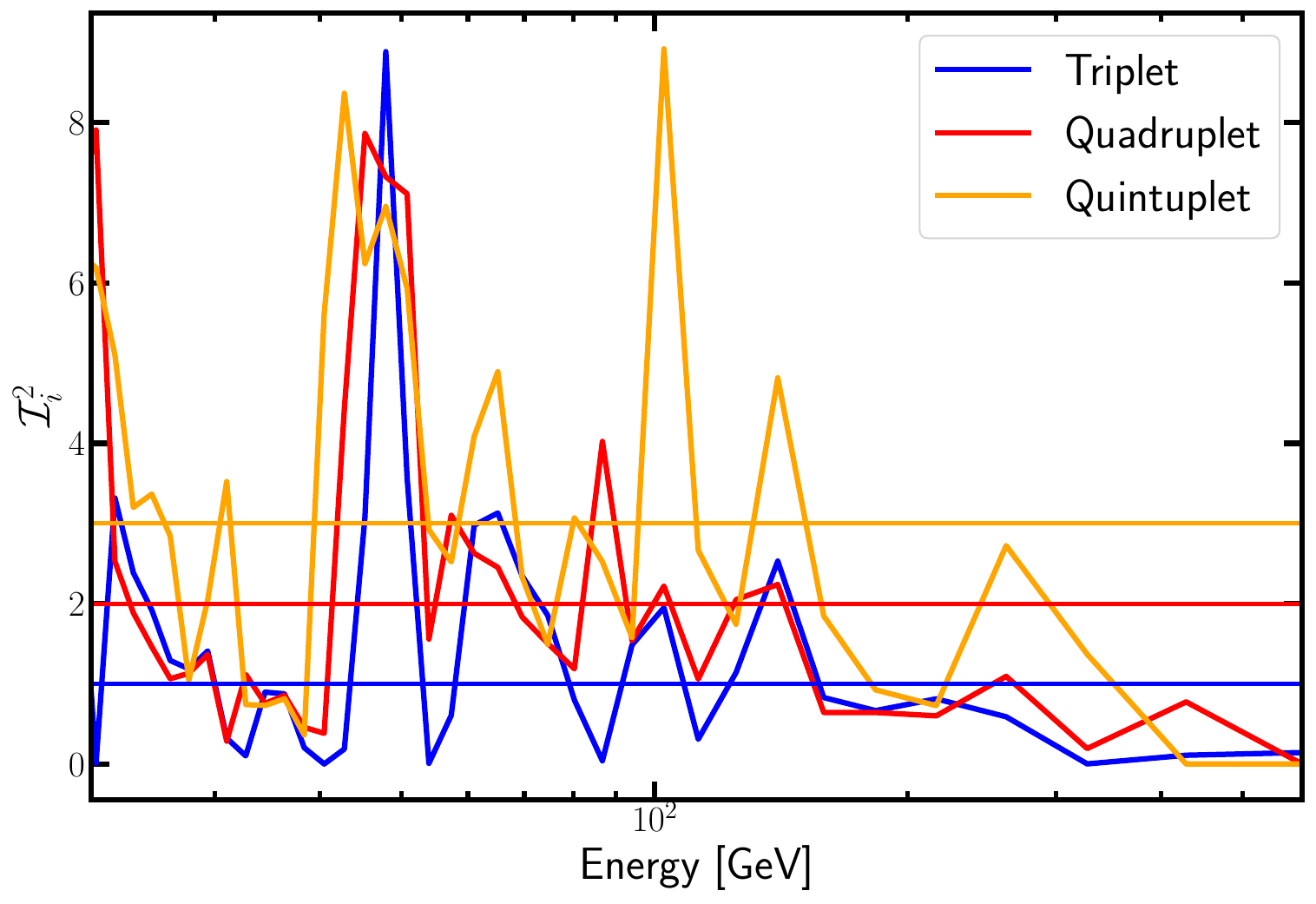}\\
\captionsetup{justification=raggedright}
\caption{The  contributions to the irregularity estimator from individual triplets, quartets, and quintets. The verticle lines represent the expected averaged estimator ($\mathcal{I}^2/d.o.f$) in the absence of irregularities.}.\label{fig: individual}
\end{figure}

Following our analysis, the AMS-02 positron spectrum exhibits more irregularities at lower energies compared to higher energies. For instance, when the energy range is restricted to above 100 GeV, the corresponding $p$-value is only 0.503, in contrast to 0.08 observed when the energy range is above 20 GeV. This result is further elucidated in Fig.~\ref{fig: individual}, where we plot the contributions to the irregularity estimator from individual triplets, quartets, and quintets. It is evident that there are more deviations from a smooth spectrum between 20 GeV to 100 GeV compared to those above 100 GeV.

This observation may appear counterintuitive, as one might naturally expect more irregularities with increasing energy. However, the strength of the irregularity estimator is not solely determined by the magnitude of the deviation from a smooth spectrum, but also by the precision of the measurement, as demonstrated by Eq.~\ref{eq:estimator}. Due to the larger statistics and consequently smaller error bars at lower energies, it is plausible that the deviation from a smooth spectrum contributed by an old and nearby pulsar is more pronounced in the measured AMS-02 spectrum.

%
\end{document}